\documentclass[12pt]{article}
\pdfoutput=1

\usepackage[nointlimits,reqno]{amsmath}
\usepackage{amssymb}
\numberwithin{equation}{section} 

\usepackage[pdftex,
 pdfproducer=TeXShop,
 pdfcreator=pdflatex]{hyperref}
 
\usepackage{cite}

\usepackage{xspace}
\usepackage{luty}

\newcommand{\T}{{T}}

\begin{document}

\begin{titlepage}

\title{Scale Invariance,  Conformality, \\
and Generalized Free Fields}

\author{Anatoly Dymarsky,$^*$ \ \ Kara Farnsworth,$^\dagger$\ \ 
Zohar Komargodski,$^\ddagger$\\
\smallskip
Markus A. Luty,$^\dagger$\ \ 
and\ \ 
Valentina Prilepina$^\dagger$}

\address{$^*$Skolkovo Institute of Science and
Technology, \\ Novaya St.\,100,
Skolkovo, Moscow Region, Russia, 143025 }
\address{$^\dagger$Physics Department, University of California Davis\\
Davis California 95616, USA}

\address{$^\ddagger$Weizmann Institute of Science, Rehovot 76100, Israel}

\begin{abstract}

\end{abstract}
This paper addresses the question of whether there are 4D Lorentz invariant
unitary quantum field theories with scale invariance 
but not conformal
invariance. 
An important loophole in the arguments of Luty-Polchinski-Rattazzi 
and  Dymarsky-Komargodski-Schwimmer-Theisen 
is that trace of the 
energy-momentum tensor $T$ could be a generalized free field.
In this paper we rule out this possibility.
The key ingredient is the observation that a unitary theory with scale but not
conformal invariance necessarily has a non-vanishing anomaly for global
scale transformations.
We show that this anomaly cannot be reproduced if $T$ is a generalized
free field unless the theory also
contains a dimension-2 scalar operator.
In the special case where 
such an operator is present it can be used to redefine
(``improve'') the energy-momentum tensor,
and we show that there is at least one energy-momentum
tensor that is not a generalized free field.
In addition, we emphasize that, in general, large momentum limits of correlation functions cannot be understood from the leading terms of the coordinate space OPE. 
This invalidates a 
recent argument by Farnsworth-Luty-Prilepina (FLP). 
Despite the invalidity of the general argument of FLP, 
some of the techniques turn out to be useful in the present context. 
\end{titlepage}

\section{Introduction}
An important general question is the symmetry structure of
the asymptotic UV and IR limits of 4D Poincar\'e invariant
unitary quantum field theory. 
It is expected on general grounds that these are described by a
renormalization group fixed point, and therefore the asymptotic theories are scale invariant.
Other more exotic possibilities include renormalization group limit cycles
or chaotic renormalization group flows.
In some cases, very concrete results concerning the symmetry group of scale 
invariant theories exist. 
In 2D, \Ref{Polchinski:1987dy} used the methods of \Ref{Zamolodchikov:1986gt} 
to give a non-perturbative argument that the only possible asymptotics 
(for unitary theories with a gapped spectrum of operators) is conformally 
invariant.
In 4D, \Ref{Luty:2012ww} showed using the methods of 
\Refs{Komargodski:2011vj,Komargodski:2011xv} 
that in perturbation theory about a conformal fixed
point the only possible asymptotics is conformally invariant.
This result was partially anticipated in earlier work by Jack and Osborn
\cite{Jack:1990eb,Osborn:1991gm} (see also \Refs{Fortin:2012hn,Baume:2014rla}).
In this paper we consider the question of whether there are non-perturbative
scale-invariant fixed points in 4D that are not conformal.

We begin with some generalities.
Given a unitary local 4D quantum field theory with a stress tensor $T^{\mu\nu}$,
a necessary and sufficient condition for conformal invariance is that there exists
a local operator $L$ such that
\beq
T^\mu{}_\mu = \Box L.
\eeq
This can be related to the more familiar condition for conformal invariance, 
$T^\mu{}_\mu = 0$, by defining an ``improved'' stress energy tensor
\beq
T'^{\mu\nu} = T^{\mu\nu} + \frac 1{3} (\d_\mu \d_\nu - \eta_{\mu\nu} \Box) L
\eeq
that is conserved and traceless.%
\footnote{We can also improve the stress tensor if there is a tensor operator
$L^{\mu\nu}$ such that $T = \d_\mu \d_\nu L^{\mu\nu}$, 
but a symmetric traceless tensor operator
of dimension $2$ is ruled out by unitarity bounds on dimensions of operators
\cite{Grinstein:2008qk}.}
We can define consistent correlation functions of $T^{\mu\nu}$ by coupling the theory 
in a coordinate-independent fashion to a background metric $g_{\mu\nu}$ and 
taking derivatives of the quantum effective action $W[g_{\mu\nu}]$ with respect to the metric. 
We can focus on the trace of the energy-momentum tensor by considering metrics of the
form
\beq
g_{\mu\nu} = \Om^2 \eta_{\mu\nu},
\eeq
and expanding about $\Om = 1$.
We therefore define the connected amplitudes 
\beq\eql{amplitudes}
A_n(x_1, \ldots, x_n) = \left.
\frac{\de}{\de \Om(x_1)} \cdots \frac{\de}{\de \Om(x_n)} 
W[\Om^2 g_{\mu\nu}] \right|_{\Om \,=\, 1}.
\eeq
At separated points this  is equal to the connected amplitude
$\avg{T(x_1) \cdots T(x_n)}$ (where $T = T^\mu{}_\mu$) 
which one can calculate, for example, with Feynman rules. 
The utility of the definition~\Eq{amplitudes} 
is that it also fixes the contact terms when some of the points coincide. 
Indeed, many such contact terms are fixed by symmetries 
(since $T$ is the trace of the stress energy tensor) 
and the definition~\eq{amplitudes} captures them correctly.

The contact terms are especially important for the momentum-space amplitudes,
since those are defined by integrals that include coincident points
\beq\eql{momentumamplitudes}
\begin{split}
(2\pi)^4 \de^4(p_1 + \cdots + p_n) & \tilde{A}_n(p_1, \ldots, p_n) 
\\
&= \myint d^d x_1\, e^{i p_1 \cdot x_1} 
\ldots \myint d^d x_{n}\, e^{i p_{n} \cdot x_{n}} 
A_n(x_1, \ldots, x_{n}).
\end{split}
\eeq
Below we will discuss $\tilde{A}_n(p_1, \ldots, p_n)$ at special kinematics where $p_i^2=0$
for all $i$.
In this case $\tilde{A}_n(p_1, \ldots, p_n)$ can be interpreted as a scattering 
amplitude for a dilaton field $\vph$ coupled to the theory via
\beq
\De S = \myint d^4 x\, \frac{\vph}{f} T + \scr{O}(\vph^2).
\eeq
More precisely, $\tilde{A}_n(p_1, \ldots, p_n)$ is the scattering amplitude
to leading order in an expansion in $1/f$, with the factors of $1/f$ stripped off.
These contributions are given by diagrams with only external dilaton lines,
and are therefore precisely given by the amplitude $\tilde{A}_n(p_1, \ldots, p_n)$.

It was shown in \Ref{Luty:2012ww} that the on-shell 4-point amplitude 
(by which we always mean $p_i^2 = 0$)
$\tilde{A}_4$ has no real intermediate states. 
This was generalized to all $\tilde{A}_n$ in \Ref{Dymarsky:2013pqa}.
This strongly relies on unitarity.
More precisely, one finds that at special kinematics there is a vanishing overlap:
\beq\eql{identity}
\bra\Psi \tilde{T}(p_1) \tilde{T}(p_2)...\tilde{T}(p_n) \ket{0} = 0~,\qquad p_i^2=0~. 
\eeq
for any state of the SFT $\bra\Psi$.
When writing the formula~\eq{identity}, we assume that the product of $\tilde T$s 
is defined with the contact terms induced by the prescription explained
above.

We first discuss the consequences of this result for perturbative field theories, 
and then in general. 
We emphasize that here ``perturbative'' means any theory near a conformal 
fixed point.
The conformal fixed point may itself be strongly coupled, but the
perturbed fixed point can be understood as a perturbative expansion
in the beta functions of the marginal couplings of the conformal
fixed point.
This includes ordinary weak-coupling perturbation theory as a special case.
In \Ref{Luty:2012ww} it was shown that in theories close to a perturbative fixed
point the contact structure implies that
\beq\eql{perturbativeTT}
\tilde{T}(p_1) \tilde{T}(p_2) = \tilde{T}(p_1 + p_2)
+ O(\be^2)~.
\eeq
This is essentially due to the fact that $\tilde{T} = O(\be)$.
See \Ref{Baume:2014rla} for a complete discussion.
We can then conclude from \Eq{identity} with $n = 2$
that matrix elements of $\tilde{T}$ have vanishing overlap with
any state, and the theory is therefore conformal.

Beyond perturbation theory,
\Ref{Dymarsky:2013pqa} argued that on-shell dilaton fields do not create physical SFT states
to leading order in $1/f$.
This means that the \emph{full $S$-matrix}
for dilaton scattering  at leading order in $1/f$ is trivial,
{\it i.e.}~all amplitudes are polynomials in the momenta.
This is a highly non-trivial constraint on the theory.
\Ref{Dymarsky:2013pqa} also pointed out that the condition for conformal invariance
is equivalent to the statement that the dilaton can be completely
decoupled (even off shell) to leading order in $1/f$
by a field redefinition of the form
\beq
\vph \to \vph + \frac{1}{f} L + O(1/f^2)
\eeq
for some dimension-2 operator $L$ ($L$ may vanish in the trivial case).
This shows that a unitary scale invariant theory that is {\it not} conformal corresponds to a theory with a trivial S-matrix for a particle $\varphi$, but, in spite of that, $\varphi$ is not manifestly  free. See~\Ref{Dymarsky:2013pqa} for the precise statement and its consequences.

There are several caveats to this statement.
First, there is a 4D unitary theory that is scale but not conformally invariant,
a free $0$-form gauge field.
This is a theory of a scalar $\Phi$ with a shift symmetry
$\Phi \mapsto \Phi + \la$, where $\la$ is independent of $x$,
where the shift symmetry  is interpreted as a gauge symmetry in the sense that 
it removes operators that are not shift invariant from the theory.
The theory is obviously scale invariant
and has a closed OPE.
It is not conformally invariant because the lowest-dimension nontrivial
operator $\d_\mu \Phi$ is not the descendant of any primary operator,
because the gauging eliminates $\Phi$ as an operator.
(This is similar to the case of free $U(1)$ gauge theory in $d \ne 4$
spacetime dimensions \cite{ElShowk:2011gz}, and in fact the $0$-form theory is
dual to a free 2-form gauge field in $d = 4$.)
This example is clearly special: the algebra of operators admits a natural extension, 
which does not affect the Hilbert space or the dynamics. 
In particular, if one restricts to local measurements in flat space, then this 
theory is indistinguishable from the ordinary free scalar field theory. 
This is why the diagnosis using the $S$-matrix does not detect this subtlety. 
There are no known counter-examples which are distinguishable from conformal 
field theories at the level of local physical measurements in flat space, 
consistently with the $S$-matrix argument above. 

Another important loophole in the general arguments of \Refs{Luty:2012ww,Dymarsky:2013pqa}
is that there is a simple  way to satisfy
the $S$-matrix vanishing theorem with a non-trivial (and a priori non-improvable) operator $T$,
namely $T$ can be a generalized free field.
This means that 
\beq\eql{TT}
\avg{T(x) T(0)} = \frac{C}{|x|^8}
\eeq
while higher correlators of $T$ with itself are given by Wick contractions
consisting of products of 2-point functions.
For example,
\beq
\begin{split}
\avg{T(x_1) T(x_2) T(x_3) T(x_4)}
&= \avg{T(x_1) T(x_2)} \avg{T(x_3) T(x_4)}
\\
&\qquad
{}+ \text{permutations} + \text{contact terms}.
\end{split}
\eeq
In this case, there are no connected contributions to the dilaton $S$-matrix,
consistent with the vanishing theorem.
In a conformal field theory, generalized free scalar field $\Phi$ with 
general dimension $\De$ can be easily ruled out.
From the 4-point function one can read off the spectrum of operators in the
OPE, and one finds that unless the dimension is $\De = \frac{d-1}{2}$ 
the energy-momentum tensor is absent.
(The case $\De = \frac{d-1}{2}$ is the free scalar fixed point.)
Since the OPE coefficients are symmetric in CFTs, one infers that
unless $\De = \frac{d-1}{2}$, $\Phi$ is absent from the OPE of $T_{\mu\nu}$ with $\Phi$, 
leading to a contradiction with translational invariance. 
This argument cannot be repeated in SFTs because the
OPE coefficients need not be symmetric.

In this paper we essentially rule out the possibility that $T$ is a generalized
free field in a 4D SFT.
Our argument 
proceeds as follows.
First, we show that any unitary theory that is not conformal has a non-vanishing
anomaly in global scale transformations.
This anomaly is then used to show that the 3-point
function of the energy-momentum tensor is nonzero at separated points. 
To show this we analyze the correlation function in momentum
space and carefully follow all the possible contact terms and their relation to the anomaly polynomial. 
The conclusion is that, unless there is an operator of dimension precisely 2, 
the 3-point function is nontrivial at separated points, and $T$ cannot
be a generalized free field. 

In the presence of an operator of dimension 2 that can mix with $T$, 
we have a weaker result: either the energy-momentum tensor can be improved to be 
conformal, or there is a at least one improvement such that $T$ is not a generalized
free field. 
The dimension-2 operator must be a singlet under all global symmetries,
and so apart from supersymmetric theories there seems to be nothing special
about having such an operator with dimension exactly equal to 2.
However, it would be nice to fill this gap and prove that none of the possible 
improved energy-momentum tensors in this situation is a nontrivial generalized free field.

\Ref{Farnsworth:2013osa} previously discussed the constraints on correlations of 
$T$ from the global scale anomaly, and several steps in the argument
above are based on that paper.
However, \Ref{Farnsworth:2013osa} erred in assuming that the large-momentum behavior of 
correlation functions of $T$ is given by the Fourier transform of the
leading part of the position space operator product expansion.
We show using examples that there is no simple correspondence between the
 momentum space correlation functions and the OPE in position space.%
\footnote{We thank A.~Schwimmer and S.~Theisen for many discussions of this point.}
This issue is also discussed in \Ref{Bzowski:2014qja}, 
which appeared while this paper was
being completed.

The outline of this paper is as follows. 
In section~2 we discuss the global anomaly and the constraints it places
on correlation functions of the  energy-momentum tensor.
This material is taken from~\Ref{Farnsworth:2013osa}.
In section~3, we use these constraints to rule out generalized free fields,
in the sense described above.
In section~4, we explain why the Fourier transform of the leading terms in the position space operator product
expansion does not give the correct large-momentum behavior of general 
correlation functions (in other words, we explain why short distance and large momentum  do not always correspond to the same physical regime). 
Section 5 contains our conclusions and a brief discussion of future directions.

\section{The Scale Anomaly}
As discussed in the introduction, we define correlation functions of the
ener\-gy-mo\-men\-tum tensor by coupling the theory to a background metric $g_{\mu\nu}$.
General covariance with respect to the background metric then enforces
the Ward identities that encode the conservation of the energy-momentum tensor.
It also fixes the contact terms, which are crucial to the argument below.

Global scale invariance na\"\i{}vely implies that 
$W[e^{2\si} g_{\mu\nu}] = W[g_{\mu\nu}]$,
where $\si$ is a real constant (independent of $x$).
However, these scale transformations can have an anomaly, whose most 
general form is constrained by general covariance to be a 
linear combination of the local dimension-4 terms:
\beq\eql{anomaly}
W[e^{2\si} g_{\mu\nu}]=W[g_{\mu\nu}]+
  \! \sigma\myint d^4 x\, \gap \sqrt{g} \left[ -a E_4
+ b \Box R + c W^2 - e R^2 + f R \wedge R \right].
\eeq
Here the invariants are respectively
the Euler density, the square of the Weyl tensor, the square of the Ricci scalar,
and the Pontryagin density.
The terms proportional to $a$, $b$, and $f$ are total derivatives, and can 
therefore be neglected.
The term proportional to $c$ vanishes on metrics of the form
$g_{\mu\nu} = \Om^2 \eta_{\mu\nu}$,
which we use to define the correlation functions of $T$ 
(see \Eq{amplitudes}).
This leaves only the term proportional to $e$, which determines the
anomaly for global scale transformations. Note that the $e$ anomaly is forbidden in conformal field theories by the Wess-Zumino consistency condition. In SFTs this anomaly is not only allowed, but, as we will see below, is necessary.

This anomaly has important consequences for the problem of scale 
{\it versus} conformal invariance. 
It follows from the $e$-anomaly that the 2-point function of $T$ is 
proportional to $e$, {\it i.e.}~$C \propto e$ in \Eq{TT}.
This can be seen as follows.
Although \Eq{TT} appears to scale as a power, it has an anomaly.
We can understand this in position space by computing 
\beq\eql{TTanomaly}
\left( x \cdot \frac{\d}{\d x} + 8 \right) \avg{T(x) T(0)} 
=  C \gap \frac{\pi^2}{96} \gap \Box^2 \de^4(x).
\eeq
The presence of the delta 
function singularity on the right-hand side can be seen by regulating the correlation function.
For example, we can make the replacement $x^2 \to x^2 + a^2$ and taking
the limit $a \to 0$ at the end of the calculation.
On the other hand, the anomaly equation \Eq{anomaly} 
implies the Ward identity
\beq
\begin{split}
&\left( \sum_{i = 1}^n x_i \cdot \frac{\d}{\d x_i}  + 4 n \right)
A_n(x_1, \ldots, x_n) 
\\
&\qquad\qquad\qquad\qquad
{}= \left. -36 e 
\frac{\d}{\d \Om(x_1)} \cdots \frac{\d}{\d \Om(x_n)} 
\myint d^4 x \, \Om^{-2} (\Box \Om)^2 \right|_{\Om \gap = \gap 1}.
\end{split}
\eeq
Comparing this with \Eq{TTanomaly} we see that $C \propto e$.
Another way to see this is to take the Fourier transform of \Eq{TT}, which 
yields a logarithmic UV divergence.
Regulating the divergence gives
\beq
\avg{\tilde{T}(p) \tilde{T}(-p)} =
\myint d^4 x\, e^{i p \cdot x}\, \avg{T(x) T(0)}
= -C \gap \frac{\pi^2}{192} \gap p^4 \ln p^2 + A p^4
\eeq
where $A$ is a subtraction constant whose value depends on the
regulator.
Comparing to the conformal anomaly again gives $C \propto e$.

The importance of this is that
if $e = 0$ then we have $\avg{T(x) T(0)} = 0$ for $x \ne 0$.
In unitary theories, this is sufficient to prove that $T \equiv 0$ as an operator, and
the theory is thus conformal.
If $e \ne 0$, it is still possible that the energy-momentum tensor can be improved
to $T = 0$ and the theory is conformal.

The anomaly \eq{anomaly} also has implications for higher correlation
functions of $T$.
These are easier to see in momentum space, where we define the
correlators using \Eqs{amplitudes} and \eq{momentumamplitudes}.
For example, for the three-point function (and constant infinitesimal  $\sigma$)~\Eq{anomaly} implies
\beq\eql{threeano}
e^{-4\si} \tilde{A}_3(e^\si p_1, e^\si p_2, e^\si p_3) 
= \tilde{A}_3(p_1, p_2, p_3) + 144 e \si 
(p_1^2 p_2^2 + p_2^2 p_3^2 + p_3^2 p_1^2).
\eeq
This requires that the 3-point function is nontrivial.
We will confront this with the hypothesis that $T$ is a generalized
free field in the following section. 
\Eq{threeano} does not immediately exclude the option that $T$ is a generalized 
free field, because the generalized free field ansatz only implies that the 
three-point function is zero at separated points, and contact terms 
contribute ``semi-local'' contributions proportional to 2-point functions
(see below).

\section{Generalized Free Fields}
We now suppose that $T$ is a generalized free field.
This means that correlation functions involving odd numbers of
$T$ operators are given purely by contact terms.
Contact terms can get contributions from terms in the Lagrangian proportional
to products of sources at the same spacetime point.
We will refer to these as Lagrangian contact terms.
Writing $\Om = 1 + \vph$, the most general Lagrangian contact terms are
\beq\eql{Lcontact}
\De\scr{L} = \vph \T + \vph^2 \scr{O}_4
+ \vph \Box \vph \scr{O}_2 + O(\vph^3),
\eeq
where $\scr{O}_4$ and $\scr{O}_2$ are scalar operators of dimension 4 and 2,
respectively.
Here $\scr{O}_4$ may be proportional to $T$ itself, or may be a
linear combination of $T$ and other dimension-4 operators, including 
$\Box\scr{O}_2$.
Possible coupling constants in \Eq{Lcontact} have been 
absorbed into the normalization of $\scr{O}_4$ and $\scr{O}_2$.
The 3-point function is then given by
\begin{align}
A_3(x_1, x_2, x_3) 
&= \left. \frac{\de}{\de \Om(x_1)} \frac{\de}{\de \Om(x_2)} \frac{\de}{\de \Om(x_3)} 
W[\Om] \right|_{\Om = 1}
\nonumber\\
\begin{split}
&= \avg{\T(x_1) \T(x_2) \T(x_3)}
\\
&\qquad {}+ \de^4(x_1 - x_2) \avg{\scr{O}_4(\sfrac 12(x_1 + x_2)) \T(x_3)}
\\
&\qquad {}+ \Box \de^4(x_1 - x_2) \avg{\scr{O}_2(\sfrac 12 (x_1 + x_2)) \T(x_3)}
\\
&\qquad {}+ \text{permutations}.
\eql{A3}
\end{split}
\end{align}
Another source for contact terms in $A_3$ is from the first term in~\Eq{A3}, namely, 
$\avg{T(x_1) T(x_2) T(x_3)}$ itself. Indeed, the OPE $T(x)T(y)$ may have delta functions. Those are precisely of the same form as the Lagrangian contact terms described above, so they do not need a separate treatment.

We now assume that $T$ is a generalized free field,
{\it i.e.}~that $\avg{T(x_1) T(x_2) T(x_3)}$ has no support when all the points are separated. Using~\Eq{A3}
we get 
\begin{align}
\begin{split}
A_3(x_1, x_2, x_3) 
&= 
c_4 \de^4(x_1 - x_2) \avg{\scr{O}_4(\sfrac 12(x_1 + x_2)) \T(x_3)}
\\
&\qquad{}
+ c_2 \Box \de^4(x_1 - x_2) \avg{\scr{O}_2(\sfrac 12 (x_1 + x_2)) \T(x_3)}
\\
&\qquad{}
+ \text{permutations},
\eql{A3simp}
\end{split}
\end{align}
where $c_4$ and $c_2$ arise as a combination of the Lagrangian contact terms
and the OPE contact terms in $\avg{T(x_1) T(x_2) T(x_3)}$.
We can choose the coefficients $c_4$ and $c_2$ to be unity by rescaling
the operators $\scr{O}_4$ and $\scr{O}_2$.

Fourier transforming the term proportional to $c_4$ we obtain an anomaly 
polynomial  proportional to $p_1^4+p_2^4+p_3^4$, which does not 
match \Eq{threeano}.
On the other hand, the term proportional to $c_2$ does have the correct
structure, so we can reproduce the correct $R^2$ anomaly only if the term involving
$\scr{O}_2$ is present. 
So our first result is that the theory must have a dimension-2 scalar operator, 
otherwise the generalized free field $T$ must vanish and the theory is conformal. 
Although the argument is not particularly complicated, the result is nontrivial. 
For example, the techniques  we used to establish this do not easily carry over to 
$d=3$, where it remains an open question whether generalized free fields 
(with or without dimension-one operators in the spectrum) can exist in the 
absence of conformal symmetry.

Let us now discuss the situation when an operator of dimension two is present. 
In this case, we can redefine $\T$ by adding an improvement term
\beq\eql{theimprovement}
\scr{L} \to \tilde{\scr{L}} = \scr{L}
-6  \xi \sqrt{g}\, R(g) \scr{O}_2
= 
\xi \left[ \vph \Box\scr{O}_2 - \vph \Box \vph \scr{O}_2 + O(\vph^3) \right].
\eeq
We then have
\begin{align}
\tilde{A}_3(x_1, x_2, x_3) 
&= \left. \frac{\de}{\de \Om(x_1)} \frac{\de}{\de \Om(x_2)} \frac{\de}{\de \Om(x_3)} 
\tilde{W}[\Om] \right|_{\Om = 1}
\\
\begin{split}
&= \avg{\tilde{T}(x_1) \tilde{T}(x_2) \tilde{T}(x_3)}
\\
&\qquad {}+  \de^4(x_1 - x_2) \avg{\scr{O}_4(\sfrac 12(x_1 + x_2)) \tilde\T(x_3)}
\\
&\qquad {}+ (1 - \xi) \Box \de^4(x_1 - x_2) \avg{\scr{O}_2(\sfrac 12 (x_1 + x_2)) \tilde\T(x_3)}
\\
&\qquad {}+ \text{permutations}.
\eql{A3new}
\end{split}
\end{align}
where
\beq
\tilde{T} = T + \xi \Box \scr{O}_2.
\eeq

We now show that there is at least one value of $\xi$ such that $\tilde{\T}$ is not a 
generalized free field.
Suppose on the contrary that $\tilde{T}$ is a generalized free field for all $\xi$.
Then
\begin{align}
\begin{split}
\tilde{A}_3(x_1, x_2, x_3) 
&= \tilde{c}_4\de^4(x_1 - x_2) \avg{\scr{O}_4(\sfrac 12(x_1 + x_2)) \tilde\T(x_3)}
\\
&\qquad{}
+ (\tilde{c}_2 - \xi) \Box \de^4(x_1 - x_2) \avg{\scr{O}_2(\sfrac 12 (x_1 + x_2)) \tilde\T(x_3)}
\\
&\qquad{}
+ \text{permutations},
\end{split}
\end{align}
where $\tilde{c}_2$ and $\tilde{c}_4$ may be different from the corresponding 
coefficients in \Eq{A3simp}. 
We now choose $\xi$ to make $\avg{\tilde\T(x_1) \scr{O}_2(x_2)} \equiv 0$.
This is always possible because the additional term is just an insertion of
\Eq{theimprovement}.
\beq
\avg{\tilde\T(x_1) \scr{O}_2(x_2)} = \avg{\T(x_1) \scr{O}_2(x_2)}
+ \xi \Box_1 \avg{\scr{O}_2(x_1) \scr{O}_2(x_2)},
\eeq
and the 2-point function $\avg{\scr{O}_2 \scr{O}_2}$ is necessarily non-vanishing
in a unitary theory.
We are then left with 
\begin{align}
\tilde{A}_3(x_1, x_2, x_3) 
&= \de^4(x_1 - x_2) \avg{\scr{O}_4(\sfrac 12(x_1 + x_2)) \tilde\T(x_3)}
+ \text{permutations},
\end{align}
and we cannot match the anomaly.
This means that it is not possible for all possible improvements of $T$ to be
generalized free fields.

It would be nice to show that in fact none of the existing energy-momentum
tensors can be a generalized free field, but this apparently requires
additional ideas beyond those discussed here.

\section{The OPE in Momentum Space}

In position space, the OPE gives the leading singularities of operator products
at short distances.
For example, for the 3-point function
\beq\eql{posnspaceOPE}
\avg{\scr{O}_a(\sfrac 12 x) \scr{O}_b(-\sfrac 12 x) \scr{O}_c(y)}
\stackrel{x\to 0}{\too}\ 
\sum_e C_{abe}(x) \avg{\scr{O}_e(0) \scr{O}_c(z)}.
\eeq
Short distances are expected to correspond to large momenta and one 
might therefore expect that the
Fourier transform of the \rhs\ gives the leading behavior of correlation functions in
momentum space:
\beq\eql{momentumspaceOPE}
\avg{\tilde{\scr{O}}_a(q + \sfrac 12 p) \tilde{\scr{O}}_b(-q + \sfrac 12 p)
\tilde{\scr{O}}_c(-p)} 
\stackrel{q \to \infty}{\too}
\sum_e \tilde{C}_{abe}(q) \avg{\tilde{\scr{O}}_e(p) \tilde{\scr{O}}_c(-p)}
\eeq
We will now show that this is not generally true.

There is a version of the momentum-space OPE that is proven to all orders 
in perturbation theory, but it involves matrix elements where all of the operators 
with small momenta are elementary fields.
(For a discussion, see \Ref{Collins:1984xc}.)
This is used for example in the theory of deep inelastic scattering,
where the composite operators are current operators, and 
elementary fields are used to create partonic quark and gluon states.
The results we obtain below are consistent with this perturbative
version of the OPE because we find in our examples that the breakdown occurs
when the dimensions of the operators with the small momentum are large.
This does not contradict the perturbative OPE, but it does invalidate
the arguments of \Ref{Farnsworth:2013osa}.

The first important general point is that the leading part of the position space OPE is only valid at separated points.
This means that it does not include the effects of contact terms in the operator
algebra, for example
\beq\eql{contactexample}
\scr{O}_a(x) \scr{O}_b(y) = \sum_c A_{abc} \de^d(x - y) \scr{O}_c(x) 
+ \cdots
\eeq
These are potentially important because they affect the momentum space
correlation functions.  By considering various examples, one can convince oneself that already such contact terms in position space can contribute various polynomials in momentum that dominate over the naive Fourier transformation of the separated position space OPE.
However, in a scale-invariant theory, contact terms are present
only if the dimensions of the operators satisfy special relations,
for example $\De_a + \De_b = \De_c + 4$ in \Eq{contactexample}.
Contact terms in the OPE are not present for generic operator dimensions and
are not the general reason why~\Eq{momentumspaceOPE} fails.

We demonstrate the main point with  
the 3-point function of primary scalar operators in a CFT.
In position space, we have
\beq
\avg{\scr{O}_1(x_1) \scr{O}_2(x_2) \scr{O}_3(x_3)}
= \frac{c_{123}}{|x_{12}|^{\De_1 + \De_2 - \De_3}
|x_{23}|^{\De_2 + \De_3 - \De_1}
|x_{31}|^{\De_3 + \De_1 - \De_2}},
\eeq
where $x_{ij} = x_i - x_j$.
Taking the limit $x_1 \to x_2$, we see that the coefficient of $\scr{O}_3$
in the $\scr{O}_1 \scr{O}_2$ OPE is
\beq
C_{123}(x_{12}) \propto \frac{c_{123}}{|x_{12}|^{\De_1 + \De_2 - \De_3}}.
\eeq
We now consider the 3-point function in momentum space
\beq
\avg{\tilde{\scr{O}}_1(q + \sfrac 12 p) \tilde{\scr{O}}_2(-q + \sfrac 12 p)
\tilde{\scr{O}}_3(-p)} 
= \myint d^4 x \, d^4 y \, e^{i q \cdot x} e^{-i p \cdot y}
\avg{\scr{O}_a(\sfrac 12 x) \scr{O}_b(-\sfrac 12 x) \scr{O}_c(y)}
\eeq
and consider the limit $q \gg p$.
We focus on the $y$ integral, given by
\beq\eql{theF}
F(p, x) = 
\myint d^d y\,
\frac{e^{-i p \cdot y}}
{|y + \sfrac 12 x|^{\De_3 - \De_{12}} |y - \sfrac 12 x|^{\De_3 + \De_{12}}},
\eeq
where $\De_{12} = \De_1 - \De_2$.
Since we are interested in $q\gg p$, we may na\"\i{}vely take $x \to 0$ and obtain
\beq\eql{naiveintegral}
F(p, 0) = \myint d^d y\, \frac{e^{-i p \cdot y}}{|y|^{2\De_3}}
\propto |p|^{2\De_3 - d}.
\eeq
If the OPE commuted with Fourier transformation,  then from~\eq{naiveintegral} we would infer that the 3-point function in the limit $q\gg p$ behaves like $q^{\Delta_1+\Delta_2-\Delta_3-d}p^{2\Delta_3-d}$.
However, when $\De_3 > \frac{d}{2}$
the integral \eq{naiveintegral} is divergent for
small $y$, and therefore the OPE  limit is more subtle.

We can see this explicitly by doing a more careful calculation.
We first combine the denominators using Feynman parameters to get
\begin{align}
F(p, x) &= \int_0^1 d\la\, \scr{N}(\la) \myint d^d y\,
\frac{e^{-i p \cdot y}}
{[y^2 + 2(\la - \frac 12) x \cdot y + \frac 14 x^2]^{\De_3}}
\nonumber\\
&= \int_0^1 d\la \, e^{i(\la - \frac 12)p \cdot x} \scr{N}(\la) \!
\myint d^d y \, \frac{e^{-i p \cdot y}} 
{[y^2 + \la (1 - \la) x^2]^{\De_3}},
\end{align}
where we shifted the $y$ integral in the last line and defined
\beq
\scr{N}(\la) = 
\frac{\Ga(\De_3) 
\la^{\frac 12 (\De_3 + \De_{12}) - 1} 
(1 - \la)^{\frac 12 (\De_3 - \De_{12}) - 1}}
{\Ga\bigl(\sfrac 12 (\De_3 + \De_{12})\bigr)
\Ga\bigl(\sfrac 12 (\De_3 - \De_{12})\bigr)}.
\eeq
Writing this in spherical coordinates 
and performing the integral over the polar angle we obtain
\beq
F(p, x) = S_{d-1} |p|^{2\De_3 - d}
\int_0^1 d\la \, \scr{N}(\la) e^{i(\la - \frac 12)p \cdot x} \!
\int_0^\infty du\, \frac{u^{d-2} \gap \sin(u)}{[u^2 + \la(1 - \la) p^2 x^2 ]^{\De_3}},
\eeq
where $u = |p| |y|$ and $S_{d-1}$ is the volume of the $(d-1)$-dimensional sphere.
The $u$ integral converges for
$d > 0$ and $\De_3 > \frac 12(d-2)$, {\it i.e.}~whenever
$\De_3$ satisfies the unitarity bound.
The $u$ integral can be written in terms of generalized hypergeometric functions.
Expanding for small $p^2 x^2$ gives
\beq\eql{expandOPE}
\begin{split}
\int_0^\infty du\, \frac{u^{d-2} \gap \sin(u)}{[u^2 + s ]^{\De_3}}
&= -\cos\bigl( \sfrac{\pi}{2} (d - 2\De_3) \left[1 + O(s) \right]
\\
&\qquad\qquad{}
+ s^{\frac{d}{2} - \De_3} \frac{\Ga(\frac{d}{2})\Ga(\De_3 - \frac{d}{2})}{2\Ga(\De_3)}
\left[ 1 + O(s) \right],
\end{split}
\eeq
where $s = \la(1-\la) p^2 x^2$.
The first (second) series dominates for $\De_3 < \frac{d}{2}$
($\De_3 > \frac{d}{2}$).
The $\la$ integral has potential singularities at $\la = 0, 1$,
but these are absent as long as
\beq\eql{De12cond}
|\De_1 - \De_2| < \begin{cases}
\De_3 & \De_3 < \frac{d}{2} \\
\frac{d}{2} & \De_3 > \frac{d}{2}.
\end{cases}
\eeq
It may be possible to extend the validity of these results beyond \Eq{De12cond}
by analytic continuation, but we will not address that here.
Note that this condition includes the case $\De_1 = \De_2$,
the case of interest in this paper.
At least in the case where \Eq{De12cond} is satisfied,
we then obtain
\beq\eql{noOPE}
\avg{\tilde{\scr{O}}_1(q + \sfrac 12 p) \tilde{\scr{O}}_2(-q + \sfrac 12 p)
\tilde{\scr{O}}_3(-p)} 
\sim \begin{cases}
q^{\De_1 + \De_2 - \De_3 - d} p^{2\De_3 - d} & \De_3 < \frac{d}{2}\\
q^{\De_1 + \De_2 + \De_3 - 2d} & \De_3 > \frac{d}{2}
\end{cases}
\eeq
We obtain the na\"\i{}vely expected OPE behavior only if $\De_3 < \frac{d}{2}$,
while for $\De_3 > \frac{d}{2}$ the leading behavior is independent
of the small momentum $p$. In the latter case, the momentum OPE regime does not correspond to the coordinate space OPE regime.
This behavior is in fact easy to understand.
For $\De_3 > \frac{d}{2}$ the $y$ integral is dominated
by $y \sim x$ rather than $y \sim p^{-1}$.
This means that the momentum-space 3-point function is not dominated by the 
regime $|x| \ll |y|$, and hence, it does not correspond to the position space OPE.

We can also obtain \Eq{noOPE} from the general momentum-space formula 
\cite{Bzowski:2013sza}
\beq\eql{CFT3ptmomentumspace}
\begin{split}
\avg{ \tilde{\scr{O}}_1(p_1)  & \tilde{\scr{O}}_2(p_2)  \tilde{\scr{O}}_3(p_3)}
\propto \gap |p_1|^{\De_1 - \frac{d}{2}} |p_2|^{\De_2 - \frac{d}{2}} |p_3|^{\De_3 - \frac{d}{2}} 
\\
&\ \ \ \ {} \times
\int_0^\infty \! dx\,
x^{\frac{d}{2} - 1} \gap
K_{\De_1 - \frac{d}{2}}(|p_1| x) K_{\De_2 - \frac{d}{2}}(|p_2| x) K_{\De_3 - \frac{d}{2}}(|p_3| x)
\,,
\end{split}
\eeq
using the asymptotic expansion of the Bessel functions.

In general, Fourier transforms for high-dimension operators must be defined by analytic continuation, but as we have seen, in spite of this, 
the short distance and high momentum limits may fail to be identical. 

Let us consider one additional instructive example. Define the operator $\scr{O} = \Phi^2$ in free scalar field theory.
Since the dimension of $\Phi^2$ is $d - 2$, it follows from the arguments above
that in the limit $q\gg p$ we have 
\beq\eql{phi2OPE}
\avg{\tilde{\scr{O}}(q + \sfrac 12 p) \tilde{\scr{O}}(-q + \sfrac 12 p) \tilde{\scr{O}}(-p)}
\sim \begin{cases}
\displaystyle\vphantom{\Biggl\{} \frac{p^{d-4}}{q^2} & d < 4 \\
q^{d- 6} & d > 4
\end{cases}
\eeq
Using dimensional regularization, the momentum-space 3-point function is
\beq
\avg{\tilde{\scr{O}}(p_1) \tilde{\scr{O}}(p_2) \tilde{\scr{O}}(p_3)}
\propto \myint d^d \ell\, \frac{1}{\ell^2 (\ell + p_1)^2 (\ell - p_3)^2}
\propto \int_0^1 dy \int_0^{1-y} dx\, 
(M^2)^{\frac{d}{2} - 3}
\eeq
where $x$ and $y$ are Feynman parameters and
\beq
M^2 = [ x(1 - x) - 2 x y ] p_1^2 + 2 x y p_2^2
+ [ y(1 - y) - 2 x y] p_3^2.
\eeq
For $d > 4$ the integrals over the Feynman parameters are convergent, and we
immediately obtain the result in the second line of \Eq{phi2OPE}.
For $d \le 4$ we must be more careful in evaluating the Feynman parameter
integrals.
For $d = 4$, the $x$ integral converges only for $y \ne 0$, and we obtain in the momentum space OPE limit
\beq
\avg{\tilde{\scr{O}}(q + \sfrac 12 p) \tilde{\scr{O}}(-q + \sfrac 12 p) \tilde{\scr{O}}(-p)}
\propto \frac{1}{q^2} \ln q^2 
\qquad (d = 4).
\eeq
For $d = 3$, the momentum integral is convergent and can be performed directly
without Feynman parameters, yielding the result in \Eq{phi2OPE}.

Finally, let us re-consider the example discussed in \Ref{Farnsworth:2013osa}, 
where we minimally couple a scalar field in four dimensions to a conformal 
background metric. 
Writing  
the background metric as
$g_{\mu\nu} = \Om^2 \eta_{\mu\nu}$
and expanding about $\Om = 1$ (as before, we denote $\vph = \Om - 1$), we have
\beq\eql{Lag}
\scr{L} = \sfrac 12  \sqrt{g}\, g^{\mu\nu} \d_\mu \Phi \d_\nu \Phi
= \sfrac 12 (1 + \vph)^2 (\d\Phi)^2.
\eeq
One may also add a curvature coupling, but, for simplicity, we do not do that. In the context that $\Phi$ is the zero-form gauge field (i.e. it has a gauged shift symmetry) then such a coupling to curvature is forbidden. In the case that we do not add a coupling to curvature, there are no dimension 2 operators in the Lagrangian~\eq{Lag} (there are also no dimension-2 operators in the algebra of operators $(\d\Phi)^2(x)(\d\Phi)^2(y)$).
The 3-point function of the dilaton $\vph$ has two terms, one proportional to
the 3-point function of $(\d\Phi)^2$ and one proportional to its 2-point function.
The contribution to the dilaton 3-point function from the 2-point
function of $\scr{O}$ has the dependence
\beq
\avg{\tilde{\scr{O}}(p_1) \tilde{\scr{O}}(-p_1) } + \text{permutations}
\propto (p_1^4 + p_2^4 + p_3^4) \left[ A + B \ln \frac{p^2}{\mu^2} \right].
\eeq
The scale anomaly requires that the coefficient of the logarithm in the dilaton
3-point function be proportional
to $p_1^2 p_2^2 + \text{permutations}$, hence, this must come from the three-point function of $(\d\Phi)^2$. The 3-point function of $\scr{O} = (\d\Phi)^2$ in position space is given by
\beq\eql{dilaton}
\avg{\scr{O}(x_1) \scr{O}(x_2) \scr{O}(x_3)}
\propto \frac{1}{x_{12}^4 x_{23}^4 x_{31}^4}.
\eeq
The previous  discussion therefore applies. One can show by an explicit computation that the correct anomaly polynomial is reproduced from the Fourier transform of~\Eq{dilaton}. 

\section{Conclusions}
Motivated by the constraints on SFTs of \Refs{Luty:2012ww,Dymarsky:2013pqa}, 
we have considered the possibility that the trace of the energy-momentum tensor
$T$ is a generalized free field.
Using some techniques from \Ref{Farnsworth:2013osa} we have been able to essentially
rule this out.
The only possibility that remains is the case where the theory contains
a singlet scalar operator $L$ of dimension precisely 2.
In this case we are able to show that there is an improvement of
$T$ that is not a generalized free field. We have not shown that a generalized free field must be altogether absent in this case. 
We hope that progress on this can be made in the future.

There are many open directions for further work. 
One question is whether these ideas can be extended to other dimensions.
Like previous constraints on 2D and 4D SFTs, the arguments of the present paper
make heavy use of various anomalies, which are absent in odd dimensions.
Straightforward attempts to extend the ideas of 
\Refs{Komargodski:2011vj,Komargodski:2011xv,Luty:2012ww,Dymarsky:2013pqa} 
to 6D fail for other reasons (see \Ref{Elvang:2012st}), 
and it appears that new ideas are needed in that case also.
In holographic theories in both even and odd dimensions, there is a
unified understanding of $c$-theorems defined using entanglement entropy
\cite{Girardello:1998pd,Freedman:1999gp,Casini:2004bw,
Myers:2010tj,Casini:2011kv,Casini:2012ei}.
These are only partially understood from the purely field theory 
perspective, and further work along these lines may shed light on the
relation between scale and conformal invariance as well.
Finally, it would be very interesting to study the space of supergravity solutions describing low energy compactifications of string/M theory to see whether there exist geometries that can be interpreted as holographic duals of scale invariant field theories that are 
not conformal.
Some steps in this direction were taken in \Ref{Nakayama:2009fe}.

\section*{Acknowledgments}
We are very grateful to R. Rattazzi, A. Schwimmer, S. Theisen
for useful discussions.
A.D. is supported by the grant RFBR 12-01-00482.
Z.K.  is supported by the ERC STG grant number 335182, 
by the Israel Science Foundation under grant number 884/11, 
by the United States-Israel Binational Science Foundation (BSF) under 
grant number 2010/629, 
and by the I-CORE Program of the 
Planning and Budgeting Committee and by the Israel Science Foundation under 
grant number 1937/12.  
K.F., M.A.L., and V.P. are supported by the US Department of Energy under
grant DE-FG02-91ER40674.
Any opinions, findings, and conclusions or recommendations expressed in this material 
are those of the authors and do not necessarily reflect the views of the funding agencies. 

\newpage
\bibliographystyle{utphys}
\bibliography{mycites}

\end{document}